\documentclass[USletter,11pt,twoside]{article}

\usepackage[utf8]{inputenc}
\usepackage[english]{babel}
\usepackage[T1]{fontenc}
\usepackage{url}
\usepackage[colorlinks=true, allcolors=blue]{hyperref}
\usepackage{graphicx}
\usepackage[left=85pt, right=85pt, top=75pt, bottom=85pt,
headheight=15pt,%
headsep=10pt]{geometry}%
\usepackage{amsmath, amssymb}
\usepackage{authblk}
\usepackage{algorithm, algpseudocode}
\usepackage{mathptmx}
\usepackage{amsmath,bm}
\usepackage{cite}
\DeclareMathOperator{\sinc}{sinc}

\usepackage[figurename=Fig.]{caption}
\DeclareSymbolFont{matha}{OML}{txmi}{m}{it}
\DeclareMathSymbol{\varv}{\mathord}{matha}{118}

\title{Polarization-Controlled Nonlinear Computer-Generated Holography}

\author[1,2,*]{Lisa Ackermann}
\author[1]{Clemens Roider}
\author[1,2]{Kristian Cvecek}
\author[1,2]{Nicolas Barr\'{e}}
\author[1]{Christian Aigner}
\author[1,2]{Michael Schmidt}

\affil[1]{Institute of Photonic Technologies, Friedrich-Alexander-Universität Erlangen-N\"{u}rnberg, Konrad-Zuse-Stra\ss e 3/5, 91052 Erlangen, Germany}
\affil[2]{School of Advanced Optical Technologies (SAOT), Friedrich-Alexander-Universität Erlangen-N\"{u}rnberg, Paul-Gordan-Stra\ss e 6, 91052 Erlangen, Germany}
\affil[*]{Corresponding author: lisa.ackermann@lpt.uni-erlangen.de}

\begin{document}

\maketitle

\begin{abstract}\noindent
  Dynamic phase-only beam shaping with a liquid crystal spatial light modulator is a powerful technique for tailoring the intensity profile or wave front of a beam. While shaping and controlling the light field is a highly researched topic, dynamic nonlinear beam shaping has hardly been explored so far. One potential reason is that generating the second harmonic is a degenerate process as it mixes two fields at the same frequency. To overcome this problem, we propose the use of type~II phase matching as a control mechanism to distinguish between the two fields.
  Our experiments demonstrate that distributions of arbitrary intensity can be shaped in the frequency-converted field at the same quality as for linear beam shaping and with conversion efficiencies similar to without beam shaping. We envision this method as a milestone toward beam shaping beyond the physical limits of liquid crystal displays by facilitating dynamic phase-only beam shaping in the ultraviolet spectral range.
\end{abstract}

\section{Introduction}\label{sec:introduction}
The first operating laser in the early 1960s \cite{MAIMAN1960} was the dawn for many research fields in modern optics although some of their fundamental effects were already demonstrated or proposed in theory decades earlier. Holography and nonlinear optics emerged independently from each other but both fields benefited from the high coherence and high power of new light sources. 
Holography is based on the interference of light waves and incorporates phase and amplitude information to go beyond photography. Dynamic phase-only beam shaping with a liquid crystal spatial light modulator (LC-SLM) is a method emerging from holography for arbitrarily controlling the intensity distribution of the beam with many applications in research \cite{Maurer2010, rubinsztein2016roadmap, Accanto:18} and industry \cite{sugioka2014ultrafast, Sugioka+2017+393+413, ORAZI2021543}. As this method only modulates the wave front, there are no significant losses. As a drawback, liquid crystal displays are technically limited to the visible, near-infrared and mid-infrared spectral ranges. This is not an insurmountable problem, as frequency conversion processes such as second harmonic generation or sum frequency generation are coherent processes that preserve the phase of the impinging fundamental wave. Combining nonlinear optics and holography allows for shaping the light field at the fundamental while achieving the targeted outcome at the frequency-converted field. Even though both research fields can be combined, the concept of nonlinear holography is only currently emerging.\\%
Yariv showed decades ago that four wave mixing can be interpreted as holographic recording and reconstruction and proposed using it for the realization of real time holography \cite{Yariv1978}. In this process, the interaction between the fields can be interpreted as one field diffracted by the shaped pattern of another field. Meanwhile many investigations followed on the nonlinear conversion of structured light for the conservation of singularities \cite{abraham1990overview, Basistiy1993} and orbital or spin angular momentum and vortex beams \cite{Dholakia1996, Courtial1997, Mair2001, Shao2013, zhou2014generation, Li2015, Steinlechner2016, Buono2018, Qiu2018, Shen2019, Li2021, PinheirodaSilva2021, Hancock2021}. Here, the physical principles of the nonlinear conversion of structured light have been well-explored and several works use beam characteristics such as polarization \cite{Buono2018}, differing wavelengths \cite{zhou2014generation, Li2015, Steinlechner2016} or non-collinear geometries \cite{Qiu2018} as a control mechanism for the nonlinear conversion of optical vortices. The recent review paper published by Buono and Forbes gives an overview on nonlinear optics with structured light \cite{buono2022nonlinear}.\\%
Currently, there are two major approaches to nonlinear holography: directly structuring the nonlinear crystal or imaging the plane of an LC-SLM into the crystal. 
3D structuring of the nonlinear crystal leads to a modulation of the nonlinear susceptibility which shapes the wave front of the emerging light field. Such elements are called nonlinear photonic crystals as the modulation of the nonlinear susceptibility affects the beam generation and propagation \cite{Berger1998, Hong2014, Shapira2015, Xu2018, Wei2018, Liu2019, Liu2020}. There are demonstrations of a binary hologram in a nonlinear crystal \cite{Zhu2020}, a structured element combined with structured light \cite{TrajtenebrgMills2017, Liu2018StrMat} or plasmonic metasurfaces \cite{KerenZur2015}. Such 3D structured nonlinear crystals act as volume holograms or phased arrays and in theory give more degrees of freedom as a thin hologram. As their implementation is technically challenging, the freedom of design is so far strongly limited and furthermore only static solutions are possible. Such practical limitations motivate the consideration of thin holograms which are easier to realize.\\
In a dynamic approach, the plane of a phase-only LC-SLM is directly imaged into the nonlinear crystal \cite{Liu2017, Liu2018, Wu2020, Yao2021}. It thus acts as a thin hologram and brings the SLM and the nonlinear crystal together in a common plane, where the SLM dynamically shapes the wave front of the fundamental as if the hologram was directly structured into the crystal.
The phase mask on the SLM imprints a locally varying phase on the incoming light field which leads to a modification of the resulting wave front where the slope corresponds to wave vectors forming the angular spectrum. 
During the nonlinear interaction, all light fields which fulfill the phase matching condition interact and create a new, frequency-converted outgoing wave. 
If the incoming fields are not distinguishable, i.e. the fields are degenerate, all wave vectors will add up, resulting in a new outgoing wave that has an angular spectrum, containing all possible sums of incoming wave vectors. 
This complicates beam shaping of the outgoing, frequency-converted wave as the initially applied angular spectrum at the SLM is not conserved. 
Consequently, there needs to be a control mechanism which makes the incoming fields distinguishable. 
Current approaches for second harmonic generation \cite{Liu2017, Liu2018, Wu2020, Yao2021} are based on a degenerate type~I phase matching process and thus involve a non-collinear geometry to control the beam-shaped outcome. This imposes severe limitations in the quality and the efficiency of the achievable results. We utilize type~II phase matching and demonstrate high-quality and highly-efficient nonlinear beam shaping in the second harmonic with a LC-SLM in a collinear geometry:\\%
\begin{figure}[h!]%
	\centering
	\includegraphics[width=.5\linewidth]{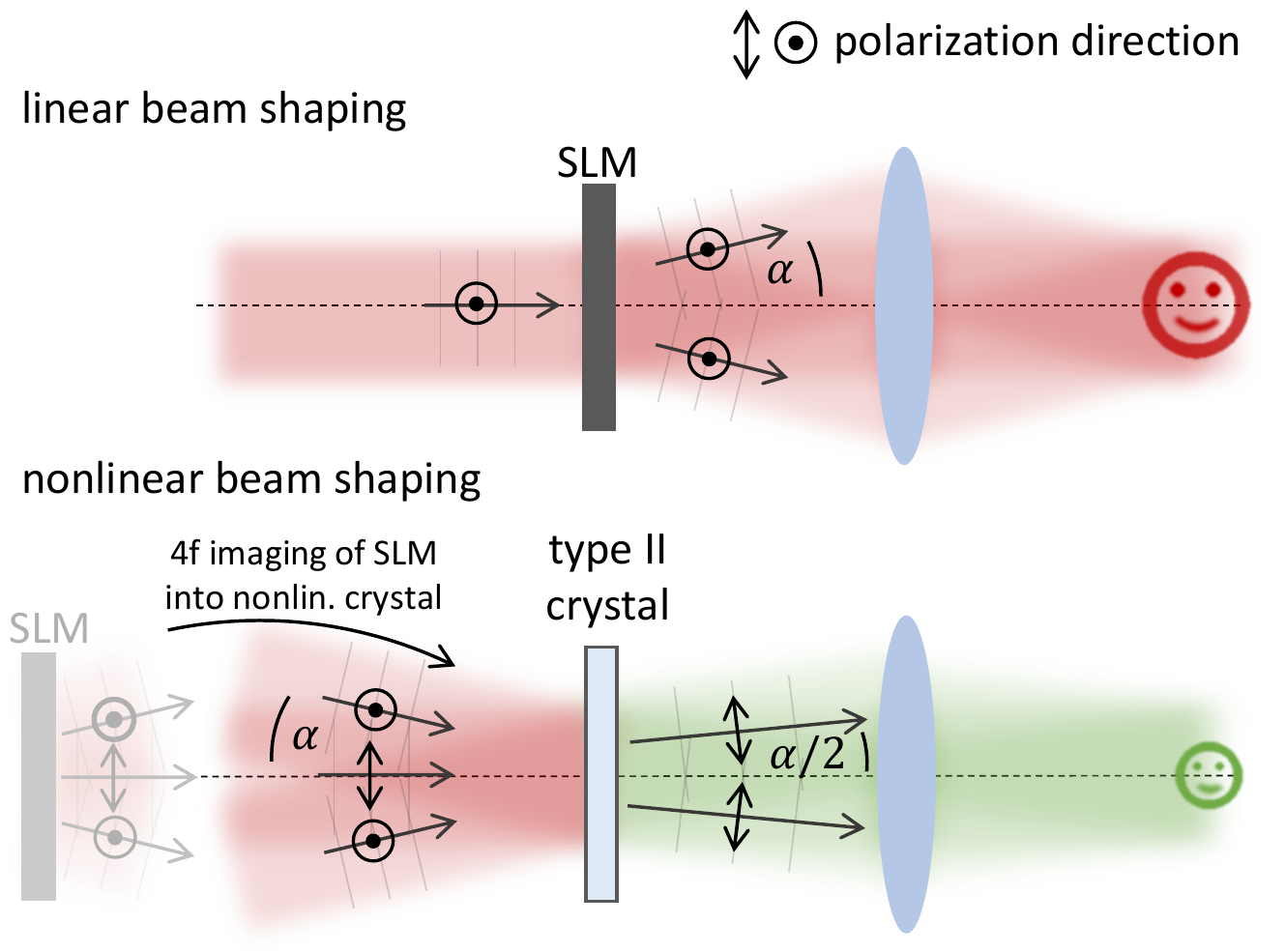}
	\caption{Basic concept of nonlinear beam shaping with linear beam shaping as reference: The applied wave front generates an angular spectrum of wave vectors which are centered around the optical axis. Their distribution results in a target image in the far field. The plane of the SLM is imaged into the nonlinear crystal. As for type~II phase matching only crossed polarization states mix efficiently, the shaped light field can only mix with the unshaped one. The addition of the wave vectors results in the tailored light field with only half the initial deflection angle $\alpha$. }\label{fig:concept}
\end{figure}%
Since the nonlinear conversion is only possible with phase matching, this constraint can be exploited for beam shaping. In type~II phase matching, only crossed polarization states can mix and through this assignment, the two fields at the fundamental are clearly distinguishable even when working in a collinear geometry. This concept is sketched in Figure~\ref{fig:concept}. By keeping one field unshaped while the other polarization carries the full phase information, there is an unambiguous assignment of the frequency-doubled wave front for every wave vector. The experimental implementation is simple as the polarization is set diagonally before it impinges on the SLM and as the SLM is polarization-sensitive, only half of the light field is shaped. After imaging this plane into the nonlinear crystal, only crossed polarization components add up for the second harmonic field. The same concept can be transferred to sum frequency generation where the differing frequencies ensure a non-degenerate mixing process. \\
Based on those physical principles, we present a setup for nonlinear beam shaping with high quality while the conversion efficiency is almost defined by the crystal's conversion efficiency without beam shaping. Experimental results are shown in Section~\ref{sec:NonlinBeamShaping}. In Section \ref{sec:MethodsModelling} we model the chosen KTP crystal to estimate the relative conversion efficiency. In Section~\ref{sec:Discussion} we discuss optimizing the conversion efficiency and give an outlook on the applicability of the method for beam shaping in the ultraviolet spectral range. Furthermore, Section~\ref{sec:Methods} presents the experimental setup.   

\section{Results}\label{sec:results}
For second harmonic generation, two fields at the fundamental frequency, $\bm{E_1}^{(\omega)}$ and $\bm{E_2}^{(\omega)}$, add up to the frequency-doubled field. From the three wave mixing process follows a quadratic relation between the fields $\bm{E}^{(2\omega)} \propto \bm{E_1}^{(\omega)} \cdot \bm{E_2}^{(\omega)}$. The phase matching condition  
\begin{align}
	\bm{k}^{(2\omega)} \stackrel{!}{=} \bm{k_1}^{(\omega)} + \bm{k_2}^{(\omega)},
	\label{eq:PhaseMatching}%
\end{align}%
requires the fundamental and the second harmonic to maintain their phase relationship during propagation through the nonlinear material to avoid destructive interference and weak conversion. 
Here $\bm{k_{1/2}}^{(\omega)}$ are the wave vectors at the fundamental frequency and $\bm{k}^{(2\omega)}$ is the resulting wave vector at the second harmonic. We first present the experimental results before we model the nonlinear crystal to evaluate the relative conversion efficiency.
\subsection{Nonlinear Beam Shaping}\label{sec:NonlinBeamShaping}
\begin{figure*}[h]%
	\centering
	\includegraphics[width=.95\textwidth]{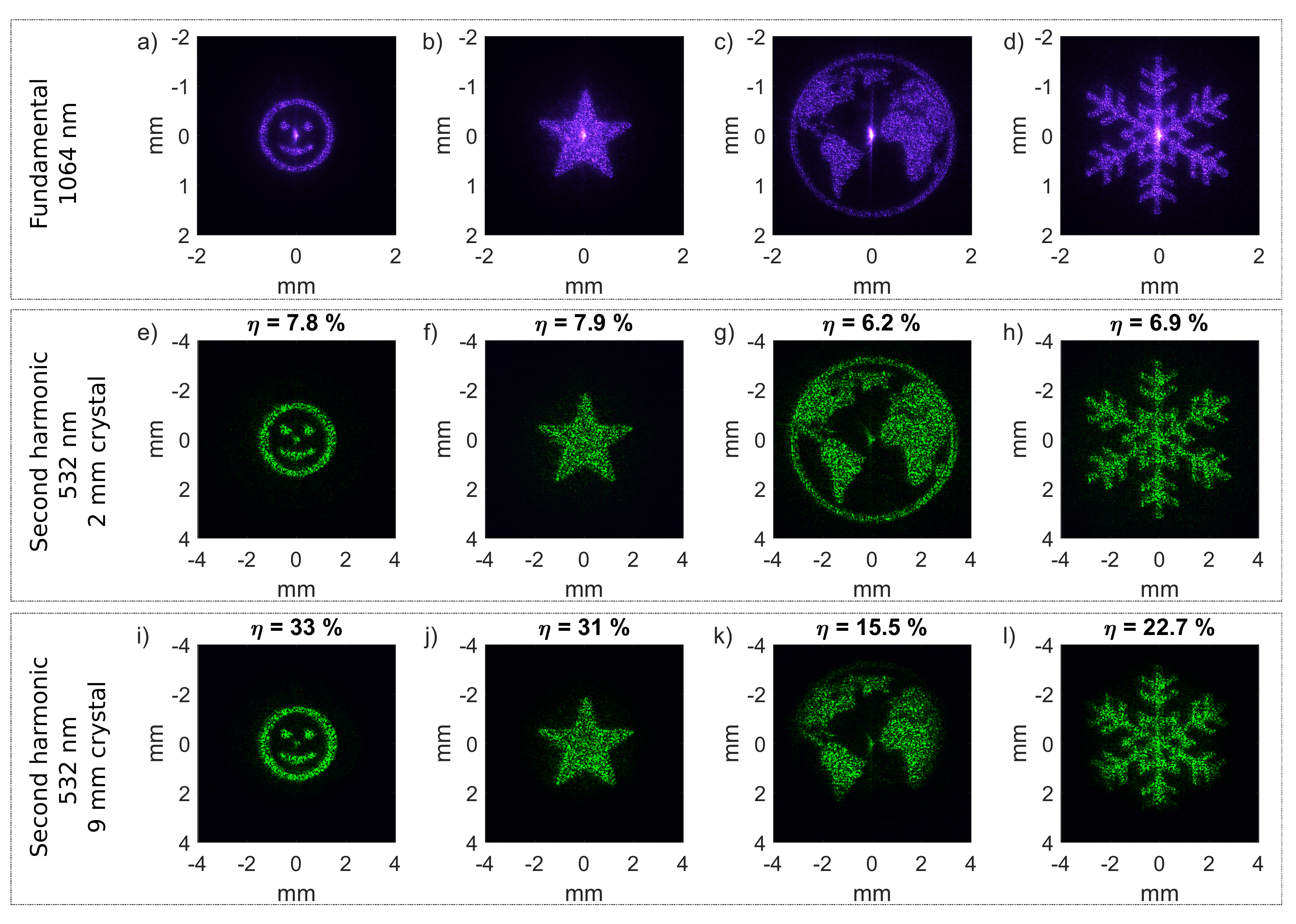}
	\caption{Experimentally recorded results for nonlinear beam shaping for two different crystal lengths and the initial result at the fundamental as reference. The measured conversion efficiency $\eta$ is denoted above the individual images. }\label{fig:Results}
\end{figure*}%
With polarization as a control mechanism, nonlinear beam shaping can be performed similar to linear beam shaping. The phase mask for wave front shaping can be calculated with the same algorithms and tools as for linear beam shaping. However, as wave front shaping generates an angular spectrum of wave vectors which are centered around the optical axis, the nonlinear crystal needs to equally support vectors which slightly deviate from the optimum phase matching angle. This required angular tolerance or acceptance is one characteristic of nonlinear crystals and it is defined for an angular deviation where the conversion efficiency drops to half of its maximum value. We chose KTP as the nonlinear crystal for our application as it exhibits a high angular acceptance. \\
Figure~\ref{fig:Results} shows experimentally recorded profiles of the second harmonic at $532\,\textrm{nm}$ and images of the fundamental at $1064\,\textrm{nm}$ (Figure~\ref{fig:Results}~a-d) as reference. The corresponding phase masks are calculated with the Gerchberg Saxton algorithm \cite{gerchberg1972practical}. When tailoring the target image in the far field, the propagation can be mathematically described by a Fourier transform. Thus, the intensity profile of the target image effectively reflects the generated angular spectrum. Its size directly shows the required angles of the wave vectors which need to be supported during frequency conversion. Consequently, the larger the targeted image size is, the higher the angular acceptance of the nonlinear KTP crystal needs to be, to equally convert the second harmonic signal for all wave vectors. \\
As a shorter nonlinear crystal exhibits a higher angular acceptance, second harmonic generation is supported for a broad spectrum of wave vectors and arbitrary target structures can be shaped over a large area. We demonstrate this in \mbox{Figure~\ref{fig:Results}~e-h} for four different target distributions with a $2\,\textrm{mm}$ long KTP crystal. All structures are homogeneously converted and the quality is comparable to the results at the fundamental. We calculate the conversion efficiency by dividing the measured power of the shaped target structure at the second harmonic by the power at the fundamental. The conversion efficiency is around $6-8\,\%$ for all structures. Conversely, a $9\,\textrm{mm}$ long KTP crystal promises higher conversion efficiency at lower angular acceptance. The results of the two smaller target structures for the $9\,\textrm{mm}$ long crystal \mbox{(Figure~\ref{fig:Results}~i,j)} are of the same quality as for the $2\,\textrm{mm}$ crystal. The conversion efficiency is \mbox{$>30\,\%$} and this is only a little less than the initial conversion efficiency of the nonlinear crystal without beam shaping which is around $40\,\%$. For the $2\,\textrm{mm}$ crystal the values are even closer with $8.5\,\%$ without and values around $6-8\,\%$ with beam shaping.
Those results demonstrate the applicability of nonlinear beam shaping in a regime of high conversion efficiency while maintaining high quality.
The homogeneous conversion in the range of the initial conversion efficiency of the nonlinear crystal is due to a plateau in the conversion efficiency for small angular deflections. We further investigate this favorable effect for beam shaping in Section~\ref{sec:MethodsModelling}. \mbox{Figure~\ref{fig:Results}~k,l} also shows the limitations of nonlinear beam shaping when working beyond this plateau. The globe and snowflake are almost cut at the borders as the required angles are not supported by phase matching. As parts of the light field are not converted, the conversion efficiency decreases. These results are shown to demonstrate the limitations outside the plateau of high conversion efficiency. It is nonetheless possible to shape a smaller target structure which is magnified with a telescope afterwards. 
\subsection{Modeling the Nonlinear Crystal}\label{sec:MethodsModelling}
\begin{figure*}[h!]%
	\centering
	\includegraphics[width=1\textwidth]{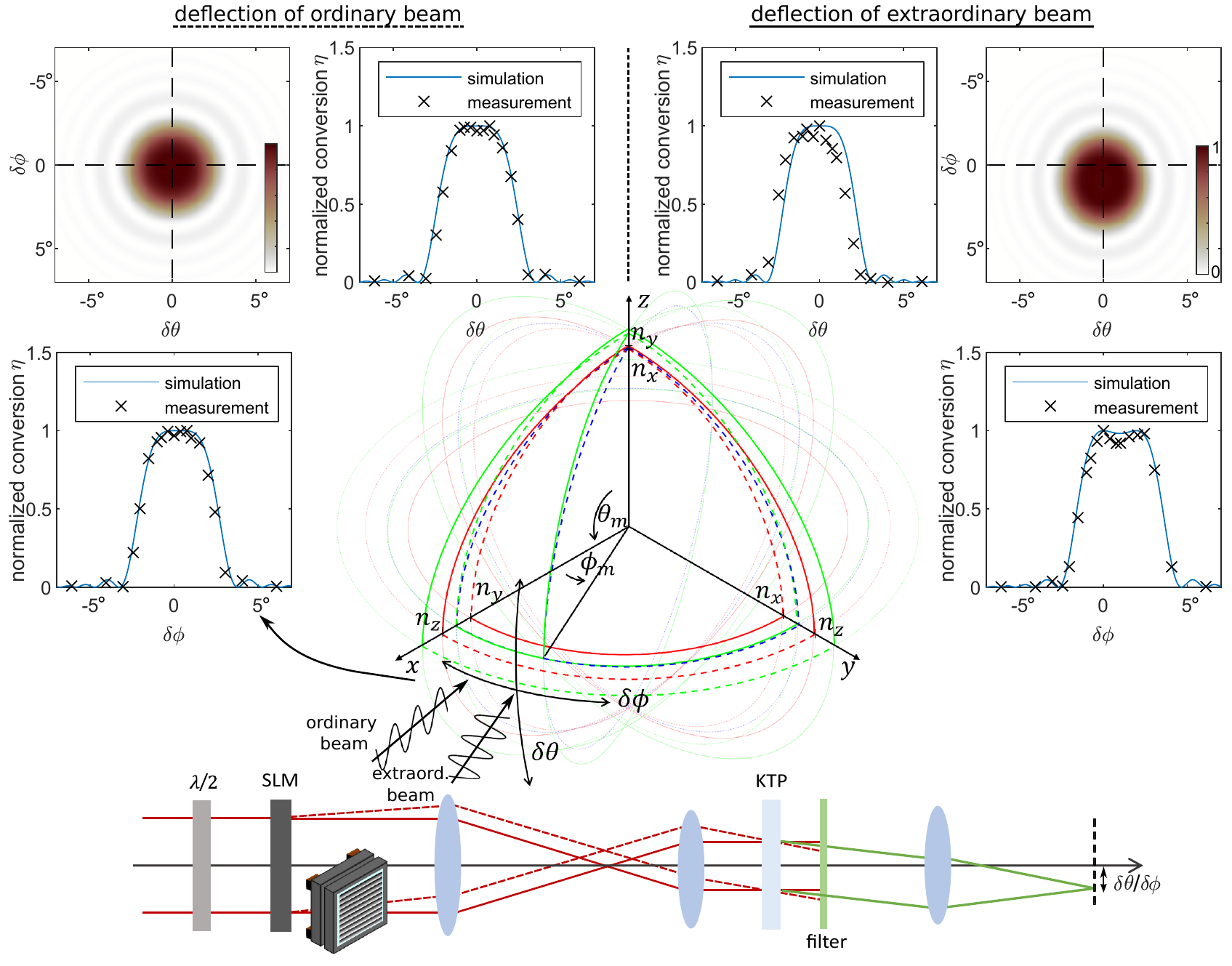}
	\caption{Normalized conversion efficiency $\eta$ for lateral deflection of the ordinary/extraordinary beam and projection along the two axes $\delta \theta$ and $\delta \phi$ for a $2\,\textrm{mm}$ KTP crystal. Experimental measurements along these axes are in good agreement with the simulated curves. The refractive index ellipsoid for the biaxial KTP crystal (dimensions not to scale) shows the phase matching angles $\theta_m = 90\,^{\circ}$ and $\phi_m = 24.8\,^{\circ}$ with the axes for beam deflection along $\delta \theta$ and $\delta \phi$. The deflection of the ordinary beam along $\delta \phi$ is indicated with two incoming fields where the extraordinary beam is polarized along the xy plane of the crystal and the deflected ordinary beam is polarized along the z-axis.}\label{fig:angularAcceptance}
\end{figure*}
The wave vector is a function of the wavelength and the corresponding refractive index, which depends on the wavelength. Consequently, the phase matching condition in Equation~\ref{eq:PhaseMatching} is generally not fulfilled if fields with different wavelengths  are propagating through a dispersive medium. Birefringence is one effect which is used for phase matching: For a biaxial crystal the refractive index is determined by three direction-dependent indices $n_x$, $n_y$, and $n_z$ along the principal axes \cite{zhang2017phase}. The effective refractive index at any position in space is given by their projection. The direction of the polarization determines the refractive index which is perceived by the corresponding light field. 
Thus, phase matching can occur under a specific angle of incidence depending on the phase matching type. Type~II phase matching is fulfilled, if two crossed polarization states mix where their average value of the refractive indices corresponds to the refractive index of the second harmonic: $(n^{o/e}(\omega) + n^{e/o}(\omega))\cdot 0.5 = n^{e/o}(2\omega)$.\\
KTP is a biaxial crystal which allows for type~I and type~II phase matching at different angles of incidence $\theta$ and $\phi$. As KTP exhibits a higher conversion efficiency for type~II phase matching, this option is the typical choice. The phase matching angles $\theta_m$ and $\phi_m$ are marked in the refractive index ellipsoid in Figure~\ref{fig:angularAcceptance}. The two red curves along the three mutually perpendicular planes of the coordinate system indicate the effective refractive index at the fundamental resulting for a field either polarized within the plane of incidence or perpendicular to that plane. From a perspective of the highest symmetry, the beam oscillating perpendicular to the plane of incidence is called the ordinary beam (dashed lines in Figure~\ref{fig:angularAcceptance}) as the refractive index is the same for any angle while a beam with the polarization axis within the plane of incidence is called the extraordinary beam (solid lines in Figure~\ref{fig:angularAcceptance}) as the effective refractive index is given by the corresponding ellipse defined by the two refractive indices of the axes spanning that plane. Likewise, the two green curves indicate the refractive index for the frequency-doubled field. As the blue curve marks the mean value of the refractive indices at the fundamental for the two polarization states, its crossing point with the green curve marks the angles for type~II phase matching. \\
Wave front shaping introduces a spectrum of wave vectors centered around the optical axis. Even in a collinear geometry, the generated angular spectrum deviates from the ideal phase matching condition. This deviation causes a refractive index mismatch and affects the conversion efficiency. Due to the high angular acceptance of the chosen KTP crystal, a deflection along $\delta \theta$ and $\delta \phi$ only causes marginal contributions to the phase mismatch. Our application for beam shaping is different from the general definition of the angular acceptance which accounts for a deflection of the total field of the fundamental: The two mixing fields have a slightly different angle as the shaped field is deflected while the other field propagates without any modifications along the phase matching angle. This results in the phase mismatch $\Delta \bm{k}$:
\begin{align}
	\begin{split}
		\Delta \bm{k} = \bm{k_1}^{\omega}(\theta_m, \phi_m) + \bm{k_2}^{\omega}(\theta_m + \delta \theta, \phi_m + \delta \phi) \\
		- \bm{k}^{(2\omega)}(\theta_m + \delta \theta /2, \phi_m + \delta \phi /2)
	\end{split}
\end{align} 
This scenario is sketched for two wave vectors in front of the refractive index ellipsoid in Figure~\ref{fig:angularAcceptance}.\\
We calculate the refractive index mismatch for a deflection along $\delta \theta$ and $\delta \phi$ by calculating the corresponding refractive indices \cite{yao1984calculations} for the wave vectors. In good approximation, the resulting wave vector at the second harmonic is generated at half of the initial deflection angle, as this is the mean value between the deflected and undeflected wave vector at the fundamental. The phase mismatch is thus given by the difference of the wave vectors projected in the direction of the frequency doubled field.
The relative conversion efficiency can be deduced from the projected mismatch $\Delta k$ by solving the following equation:
\begin{align}
	I_{2\omega} \propto I^2_\omega L^2 \sinc^2\left(\frac{\Delta k L}{2}\right)
	\label{eq:conversionEfficiency}%
\end{align}%
This analytical relation between the second harmonic intensity $I_{2\omega}$ and the intensity at the fundamental $I_{\omega}$ with respect to the crystal length $L$ and the phase mismatch $\Delta k$ follows when solving the nonlinear wave equation for the assumption of low depletion and the slowly varying amplitude approximation \cite{boyd2020nonlinear}. On this basis, we calculate the relative 2D conversion efficiency for a deviation within the plane of $\delta\theta$ and $\delta \phi$ in Figure~\ref{fig:angularAcceptance}. According to Snell's law and assuming the small angle approximation, the internal angle corresponds to the external angle connected via the effective refractive index which is perceived by the corresponding wave vector.
The graphs in Figure~\ref{fig:angularAcceptance} show the external angle, as this is the relevant value for beam shaping. \\
Besides calculating the relative conversion efficiency, we perform experiments for an angular deviation introduced by the SLM for beam shaping. To evaluate different angles, we apply a blazed grating either in horizontal or vertical direction on the SLM and image this plane into the nonlinear crystal which has a length of $2\,\textrm{mm}$. Similar to beam shaping, only one polarization component is shaped. The corresponding setup is sketched at the bottom of Figure~\ref{fig:angularAcceptance}. In the far field we measure the resulting power of the frequency-doubled field and divide it by the measured power of the field at the fundamental. This results in two curves for the relative conversion efficiency, either in the direction of $\delta \theta$ or $\delta \phi$, with respect to no deflection at $0\,^\circ$ at the optimum phase matching angle. As either the ordinary or extraordinary polarization state can be deflected within the $\delta \theta$ and $\delta \phi$ plane, two different scenarios are possible and both are shown in Figure~\ref{fig:angularAcceptance}. Deflecting the ordinary beam causes a highly symmetric profile of the relative conversion efficiency and is thus the favorable option for our application. In addition to the experimentally measured points, we add the analytically calculated curve of the relative conversion efficiency. \\
Simulation and experiment are in good agreement and show that the conversion efficiency exhibits a broad plateau. Within that range, the light is homogeneously converted, independently of the exact deflection angle. Moreover, the conversion efficiency is not significantly reduced with respect to the actual conversion efficiency without beam shaping. Thus, arbitrary intensity profiles can be shaped within that angular range without reduction in homogeneity or efficiency. The acceptance angle approximately reaches  $\pm 2\,^{\circ}$ for a $2\,\textrm{mm}$ crystal and $\pm 1\,^{\circ}$ for the $9\,\textrm{mm}$ long crystal. It makes sense to compare this value with the maximum deflection angle of the SLM which is determined by $\sin^{-1}(\lambda/(2u))\cdot 1/M$, where $u$ is the pixel pitch and $M$ is the magnification between the SLM and the crystal. The two smaller images in Figure~\ref{fig:Results} approximately acquire $19\,\%$ of the SLM's linear field of view, while the two bigger images acquire $32\,\%$. This corresponds to deflection angles of approximately $1.1\,^\circ$ and $1.9\,^\circ$ (bearing in mind that the plane of the SLM is imaged with a factor of $M=0.25$ into the nonlinear crystal and thus the maximum deflection angle is $\pm6\,^\circ$).
At the phase matching angle $\theta_{m}=90\,^\circ$, the crystal is non-critically phase-matched along $\delta \theta$. %
This means when expanding the mismatch $\Delta k$ in a Taylor series, the first derivative $\partial \Delta k / \partial \theta = 0$ due to the crystal's symmetry and thus only higher order terms with a weaker effect on the phase mismatch contribute \cite{fan1987second}. Besides the control mechanism for the wave mixing process it is crucial to choose a nonlinear crystal with a high angular acceptance. We consider temperature-controlled noncritical phase matching or quasi-phase matching as another option but do not discuss this further within the scope of this paper.

\section{Discussion}\label{sec:Discussion}
Nonlinear beam shaping is a powerful tool as it enables dynamic phase-only beam shaping in new spectral ranges. To optimize this process, two parameters should be considered: While the conversion efficiency should be maximized, the angular acceptance needs to be high enough to ensure homogeneous conversion of the angular spectrum. As both parameters are mutually dependent, we discuss the proper choice of the experimental parameters to optimize the outcome. Moreover, we give an outlook on dynamic phase-only beam shaping in the ultraviolet spectral range.

\subsection{Optimizing the Conversion Efficiency with Respect to the Angular Acceptance}
\begin{figure}[h!]%
	\centering
	\includegraphics[width=0.6\linewidth]{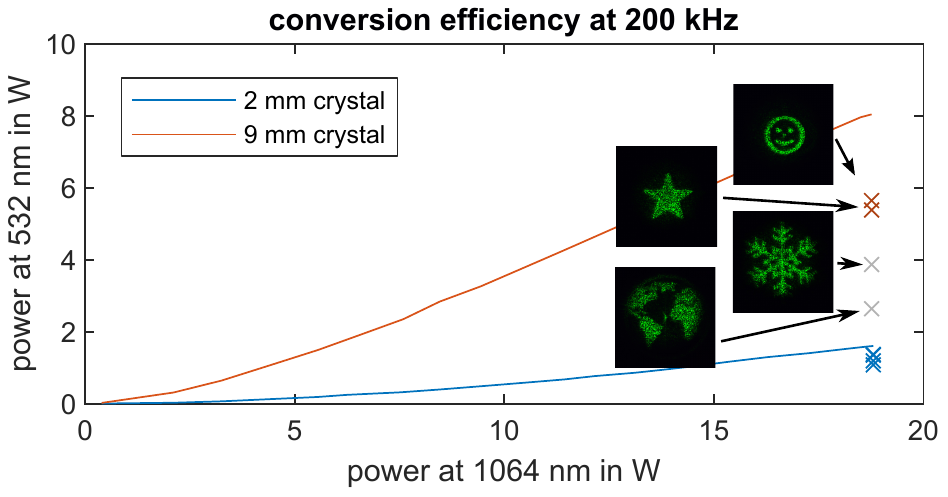}
	\caption{Measured output power vs. input power for second harmonic generation with a $2\,\textrm{mm}$ and $9\,\textrm{mm}$ KTP crystal without beam shaping. The crosses indicate the conversion efficiency of the structures in Figure~\ref{fig:Results} for beam shaping, with the two intentionally cropped structures in gray. }\label{fig:conversionEfficiency}
\end{figure}

When approaching a high conversion regime, it is more critical to maintain proper phase matching as the phase mismatch gains an increasing impact on the efficiency of the converted outcome. At high conversion, the angular range of tolerance narrows and this decreases the plateau of homogeneous conversion for the generated angular spectrum. Equation~\ref{eq:conversionEfficiency} shows the impact of the experimental parameters on the conversion efficiency while keeping the laser power constant. Both increasing the intensity and the crystal length promises higher conversion efficiency. However, both parameters also affect the angular acceptance of the nonlinear crystal.\\
The effect of the crystal length $L$ is directly stated in Equation~\ref{eq:conversionEfficiency}. While a longer crystal increases the relative conversion with $L^2$, it also increases the contribution of the phase mismatch as a multiplier of $\Delta k$ in the argument of the $\sinc$ function. We use the model of the nonlinear crystal to calculate the relative conversion of the target structures for the $9\,\textrm{mm}$ crystal as they are shown in Figure~\ref{fig:Results}. %
\begin{figure}[h!]%
	\centering
	\includegraphics[width=.55\linewidth]{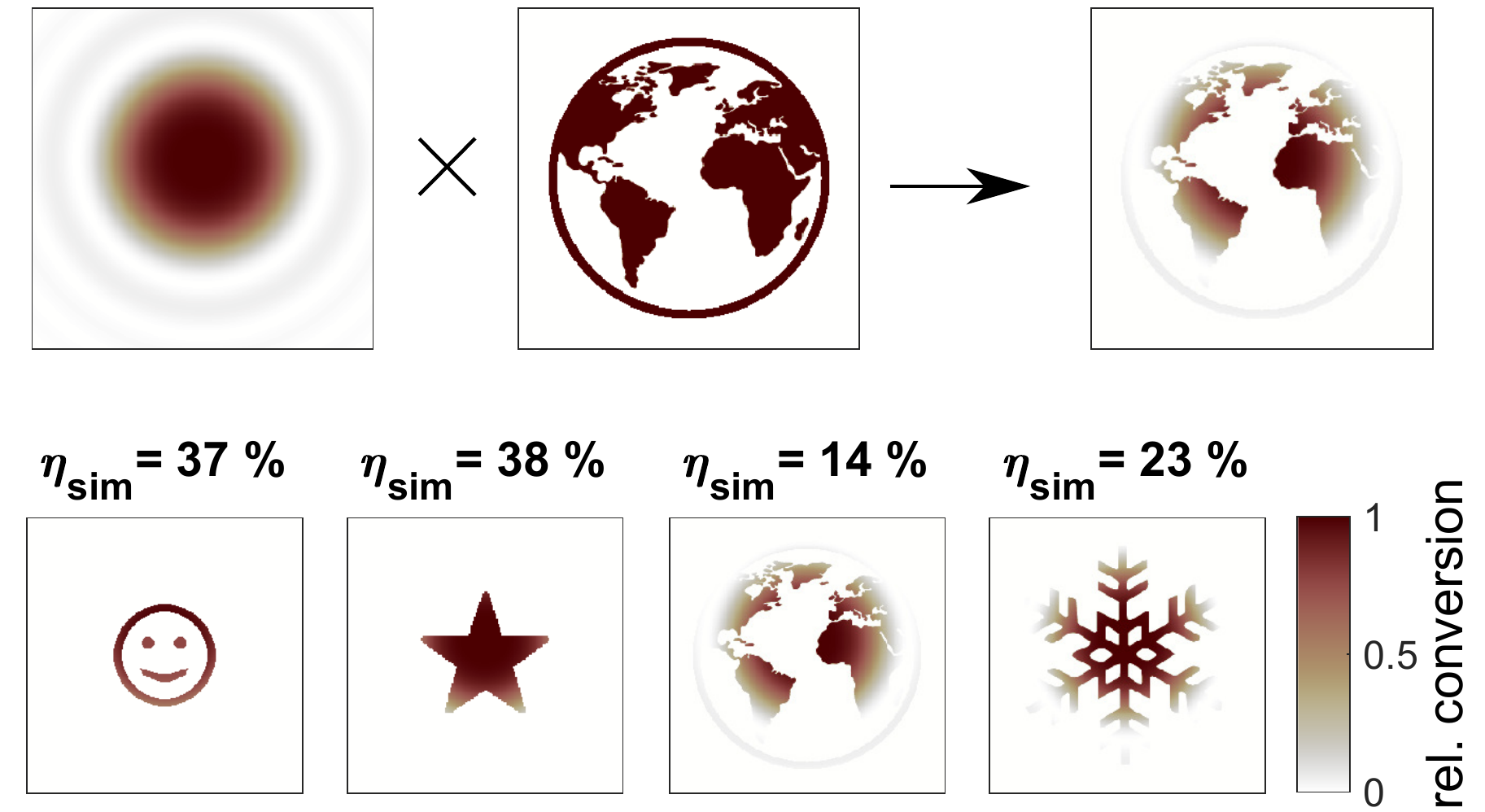}
	\caption{Simulated beam conversion for the $9\,\textrm{mm}$ crystal. The simulated efficiency is calculated from the integrated relative conversion multiplied with the efficiency of the $9\,\textrm{mm}$ crystal without beam shaping (compare Figure~\ref{fig:conversionEfficiency}).}\label{fig:sim9mm}
\end{figure}%
The beam-shaped structures directly reflect the generated angular spectrum and thus the relative conversion can be simulated when multiplying the target image with the calculated angular conversion.
Figure~\ref{fig:sim9mm} shows the simulated results with the corresponding conversion efficiency. This value is calculated as the total efficiency without beam shaping multiplied by the integrated relative conversion. Working beyond the limits of phase matching results in weak conversion and the simulated as well as the measured values for the conversion efficiency drop significantly. The experimental parameters should be chosen to work within the pleateau of homogeneous conversion as otherwise not only the quality gets worse but also the efficiency drops significantly. The comparison of the simulated and experimentally measured conversion efficiency explains the strong decrease in efficiency for the two larger structures with their measured conversion efficiency marked as gray crosses in Figure~\ref{fig:conversionEfficiency}. This decrease can be traced back to insufficient phase matching as the structures are chosen larger than the angular acceptance. 
As simulation and experiment are in good agreement, proper experimental parameters can be derived on that basis even in a high conversion regime.  \\
Similar to the crystal length $L$, an increase in intensity also improves the conversion efficiency. It can either be increased by increasing the power of the chosen light source or by reducing the illuminated area. As the laser power is typically technically limited, we will focus on the second case. Here the plane of the SLM is imaged with a certain demagnification into the nonlinear crystal to decrease the illuminated area. It is also possible to illuminate the SLM with a smaller beam but this reduces the number of illuminated pixels and unnecessarily increases the intensity on the SLM with respect to potential damage thresholds. It is thus beneficial to design the telescope with the required demagnification. This demagnification increases the intensity in the nonlinear crystal and scales the intensity with $1/M^2$ to gain higher conversion efficiencies. Likewise, the telescope increases the angular spectrum as the deflection angles are proportional to $M$. Similar to the crystal length $L$, this factor also affects the phase mismatch $\Delta k$ in the argument of the $\sinc$ function as it acts as a multiplier of the initially induced angular spectrum. Consequently, the intensity needs to be considered when modeling the phase mismatch to determine optimum parameters.\\
In our experimental setup we work with a telescope that decreases the image size by a factor of $0.25$ and thus increases the intensity by a factor of $16$. This reduction affects the deflection angles and increases them by a factor of $4$ as the effective pixel size of the SLM changes. We consider that adjusting the intensity with respect to the required angular spectrum for a given crystal length is reasonable. The resulting plateau should be exactly within the required angular range - neither broader nor narrower - to ensure homogeneous conversion with maximized efficiency.

\subsection{Towards Dynamic Phase-only Beam Shaping in UV Range}\label{sec:Outlook}
Dynamic beam shaping in the ultraviolet (UV) spectral range is highly limited as many devices absorb the UV light, including liquid crystal displays. This paper demonstrates nonlinear beam shaping from the infrared to visible spectral range. Nonetheless, other conversion processes are possible, reaching from second harmonic generation at different fundamental frequencies up to other nonlinear processes like sum frequency generation. Currently, there is research on new materials with tunable birefringence in the UV spectral range for spatial light modulators with results showing light modulation at $303\,\textrm{nm}$ \cite{Xu2023}. We see the potential of our method for approaching even deeper UV spectral ranges.
This section presents a few thoughts on proper nonlinear crystals and conversion processes for dynamic phase-only beam shaping in the UV range. \\
Ultrashort pulsed laser systems often feature nonlinear crystals for frequency conversion, for example to the second or third harmonic. High-energetic short wavelengths in the UV range of ultrashort pulsed laser systems are thus often generated with nonlinear frequency conversion from the visible or infrared range. Consequently, the initial conditions often enable a direct integration of nonlinear beam shaping into the laser setup and ideally this can be performed with conversion efficiencies close to the initial values.\\
To achieve high-quality results, the nonlinear crystal should have a broad angular acceptance and the conversion process needs to be non-degenerate. Sum frequency generation with an LBO crystal from $1064\,\textrm{nm}$ and $532\,\textrm{nm}$ to $355\,\textrm{nm}$ is one option. The major benefit of this crystal is the high angular acceptance. Similarly, sum frequency generation with CLBO \cite{Mori1995} from $1064\,\textrm{nm}$ and $266\,\textrm{nm}$ to $213\,\textrm{nm}$ should be possible. In both options the process is inherently non-degenerate as two different wavelengths mix. While the SLM can either shape the light field in the infrared or visible spectral range, only the infrared light can be shaped in the latter case as the SLM can not shape light at $266\,\textrm{nm}$. With one of the options suggested here or other nonlinear processes, we see the potential of using nonlinear beam shaping to approach the UV spectral range. This enables dynamic phase-only beam shaping beyond the physical limits of the liquid crystal display. 

\section{Materials and Methods}\label{sec:Methods}
\begin{figure}[h!]%
	\centering
	\includegraphics[width=.6\linewidth]{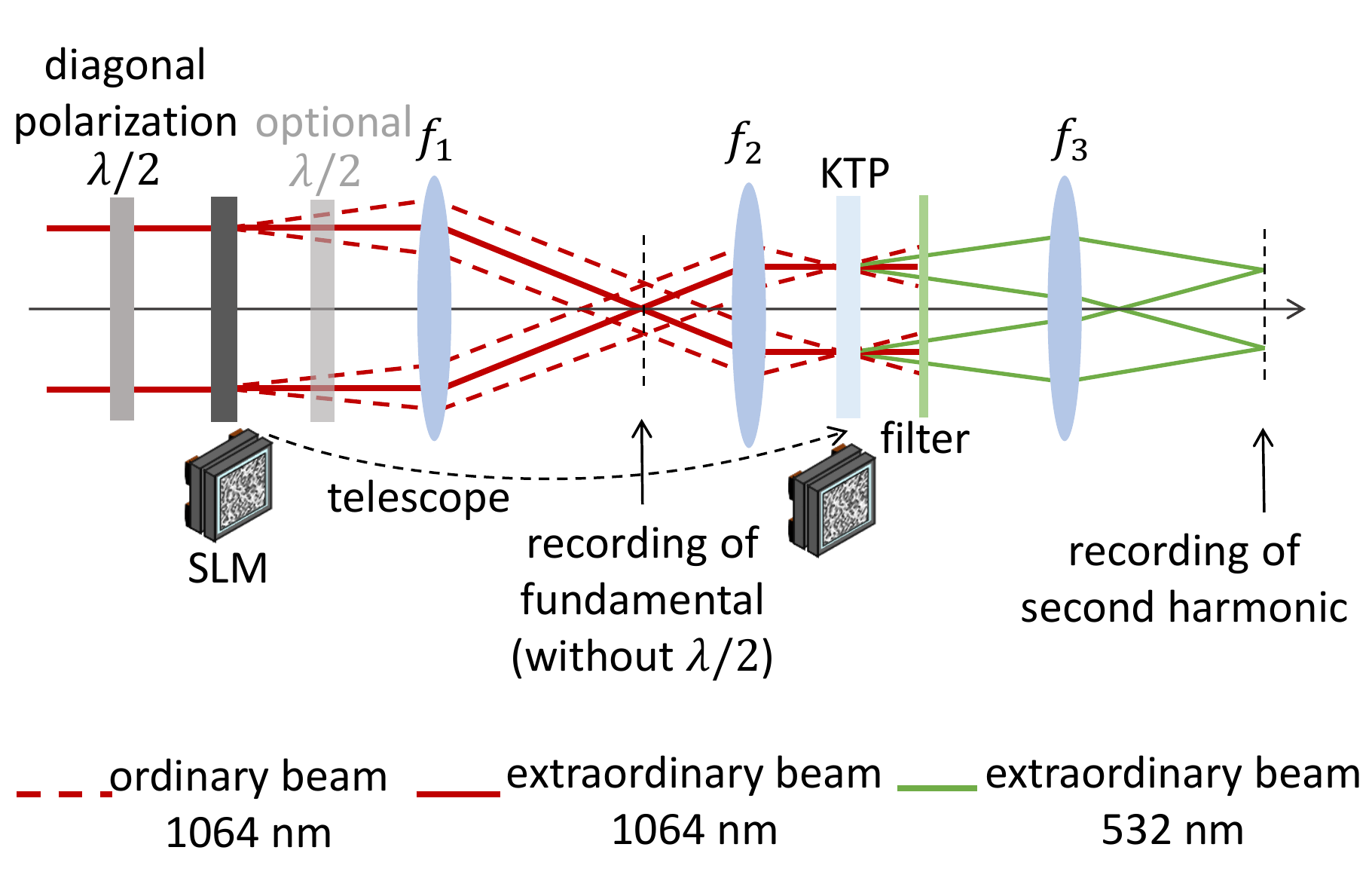}
	\caption{Experimental setup for nonlinear beam shaping: A half wave plate introduces diagonal polarization on the beam impinging on the SLM. Thus, only half of the field is shaped while the crossed polarization component remains unshaped. A telescope images the plane of the SLM into the nonlinear KTP crystal. Here, the two polarization components mix and shape the targeted field. After a filter separates the second harmonic from the fundamental, a lens images the target image, where the second harmonic signal is recorded. As the shaped result is sensitive to the deflected polarization component, a half wave plate after the SLM allows for changing the deflected and non-deflected polarization component.}\label{fig:SetupMethods}
\end{figure}%
\noindent
Nonlinear phase-only beam shaping with a liquid crystal display is possible as the applied phase information is conserved during the frequency conversion. However, as second harmonic generation is a degenerate process, the mixing waves need to be controlled by making the interacting fields distinguishable. We use the requirements of type~II phase matching to unambiguously define the outcome of the second harmonic. Under this condition, only fields of crossed polarization can mix efficiently. A half wave plate, abbreviated with $\lambda/2$ in Figure~\ref{fig:SetupMethods} is placed in front of the SLM to generate diagonal polarization with respect to the SLM's orientation. As the liquid crystals are polarization-sensitive, only half of the field is shaped, while the other part remains unshaped. Consequently, a shaped component can only mix with the unshaped plane wave front. In Figure~\ref{fig:SetupMethods} the two polarization components are indicated as the ordinary and the extraordinary beam. The polarization configuration in the figure refers to where the ordinary beam is shaped. If the extraordinary beam is shaped, an additional half wave plate after the SLM helps to rotate the polarization states correspondingly. This might be beneficial when working off-axis to separate non-diffracted light as the area of quasi-homogeneous conversion is shifted away from the center of the optical axis. 
A telescope images the plane of the SLM into the nonlinear crystal. This ensures a homogeneous intensity distribution within the nonlinear crystal which is required to convert all field components equally within a range of proper phase matching. In good approximation, the resulting wave vectors at the second harmonic reduce to half of the initially set angle of the shaped field component, as this is the sum of the two involved wave vectors. \\
The SLM is the model LSH0701010 from Hamamatsu (Hamamatsu, Japan) and has a pixel resolution of $800\,\textrm{px}\times600\,\textrm{px}$ with a pixel pitch of $20$\,µm. The beam diameter on the LC-SLM is $14\,\textrm{mm}$. 
To increase the intensity to obtain a higher conversion efficiency, we design our telescope to demagnify the beam by $M=0.25$. The corresponding focal lengths are $f_1 = 200\,\textrm{mm}$ and $f_2 = 50\,\textrm{mm}$. Before the target distribution is imaged to the far field with a $f_3 = 200\,\textrm{mm}$ lens, the remaining IR light is blocked with the filter FL532-10 from Thorlabs (NJ, USA). The laser system Fuego from Time-Bandwidth Products (CA, USA) emits $10\,\textrm{ps}$ pulses at a repetition frequency of $200\,\textrm{kHz}$ at a wavelength of $1064\,\textrm{nm}$. The power measurements were done with the power meter PM10 from Coherent (CA, USA) and PM160 from Thorlabs. We record RGB images with the camera model \mbox{UI-3000SE-C-HQ} from IDS (Obersulm, Germany).

\section{Conclusion}\label{sec13}
Nonlinear computer-generated holography is gaining increasing attention and there is a lot of research on structuring nonlinear crystals but this process is technically challenging and static. These deficiencies can be overcome by thin holograms generated with dynamic liquid crystal displays. However, research on nonlinear beam shaping is still in its infancy. 
We present a high-quality and highly efficient method for nonlinear dynamic phase-only beam shaping with a simple concept: The mixing process needs to be non-degenerate by employing a control mechanism and the angular acceptance of the nonlinear crystal needs to support the angular spectrum generated by wave front shaping. Based on this, we demonstrate beam shaping at the second harmonic with conversion efficiencies close to the ones without beam shaping. This is mainly due to the fact that we work at a plateau of constant conversion efficiency for the generated angular spectrum. While working within that range, the conversion efficiency is mainly given by the conversion efficiency without beam shaping and thus this method is highly efficient. Furthermore, the quality of the frequency-converted result is of the same quality as for beam shaping at the fundamental. We see high potential in this approach as it not only enables highly-efficient and high-quality beam shaping with thin holograms but due to its simplicity, it can be easily combined with other elaborated processes. Based on nonlinear beam shaping, the spectral range of dynamic liquid crystal displays can be extended beyond the physical limits as we discuss here for the UV spectral range.  

\section*{Funding}
The authors gratefully acknowledge funding from the Erlangen Graduate School in Advanced Optical Technologies (SAOT) by the German Research Foundation (DFG) in the framework of the German excellence initiative.

\section*{Authors' contributions}
LA and CR conceived the method. LA recorded the experimental data and performed simulations. LA, CR and KC analyzed the data. CA performed provisional experiments. LA, CR, KC, NB and MS contributed to the manuscript and MS acquired funding.

\clearpage

\let\OLDthebibliography\thebibliography
\renewcommand\thebibliography[1]{
  \OLDthebibliography{#1}
  \setlength{\parskip}{1ex}
  \setlength{\itemsep}{0pt plus 1ex}
}

\bibliographystyle{naturemag}
\bibliography{NonlinearHolographyArxiv}

\clearpage


\end{document}